\documentclass{article}
\usepackage{frascatiphys}
\usepackage{graphicx}
\usepackage{amsmath}
\begin{document}
\title{ 
Recent CLFV searches at Belle  and Belle II
}
\author{
Vindhyawasini Prasad      \\
(On behalf of the Belle II Collaboration) \\
{\em College of Physics, Jilin University, Changchun 130012, China} \\
     }
\maketitle
\baselineskip=11.6pt

\begin{abstract}
Charged lepton flavor violation (CLFV) is forbidden in the Standard Model (SM) due to massless neutrinos and absence of right-handed neutrinos. The discovery of neutrino oscillations established that the neutrinos are massive and the lepton flavor is not an exact symmetry of nature. A minimal extension of the SM that incorporates right-handed neutrinos gives rise to CLFV through loop-level diagrams involving neutrinos. However, such processes are extremely rare, for instance, the branching fraction of $\mu \to e\gamma$ is expected to be at the level of $10^{-54}$, far beyond the sensitivity of current collider experiments. In contrast, several new physics models predict CLFV decay rates within the sensitivity of current experiments, such as the Belle and Belle II experiments, which have collected the world's largest  asymmetric $e^+e^-$ collision data  near the $\Upsilon(4S)$ resonance.  The large data sample has utilized to explore the possibility of CLFV decays  of $\tau$ leptons, $B$ mesons, and bottomonium states. The report reviews some latest results on these CLFV decays at Belle and Belle II.
\end{abstract}
\baselineskip=14pt

\section{Introduction}
In the Standard Model (SM), lepton flavor is accidentally conserved due to massless neutrinos and the absence of right-handed neutrinos. The observation of neutrino oscillations~\cite{nuosc} implies that neutrinos have nonzero masses and that lepton flavor is not conserved in the neutral lepton sector. Extending the SM minimally by incorporating right-handed neutrinos gives rise to charged lepton flavor violation (CLFV) processes through loop-level diagrams involving neutrinos. These processes are, however, strongly suppressed by the squared ratio of the neutrino mass to the $W$-boson mass through the Glashow-Iliopoulos-Maiani (GIM) mechanism~\cite{gim}, resulting in extremely small branching fractions. For instance, the predicted branching fraction for $\mu \to e\gamma$ is of order $10^{-54}$~\cite{Petcov}, which is far below the sensitivity of current experiments.  Therefore, the observation of CLFV at experimentally accessible rates would constitute unambiguous evidence for physics beyond the SM.

Many extensions of the SM, including supersymmetric models~\cite{Brignole}, leptoquark models~\cite{Calibbi}, heavy-neutrino scenarios~\cite{Ellis}, and other new-physics (NP) frameworks, can enhance CLFV decay rates, with branching fractions reaching as large as $10^{-10}$--$10^{-7}$ in favorable to the current experimental sensitivity~\cite{Hisano}. An effective Lagrangian describing NP effects can be expressed as the sum of dipole terms, four-fermion interactions, and gluonic interaction terms~\cite{Hazard}. The Wilson coefficients associated with the NP operators can be constrained through global fits to measurements of processes involving CLFV interactions~\cite{Hazard}. Such NP scenarios can be explored through searches for CLFV decays at current and next-generation experiments, including the Belle and  Belle II~\cite{BelleII}, which operate as  B factory experiments.

Belle II is the upgraded successor to the Belle experiment~\cite{Belle}  and operates at the asymmetric-energy $e^+e^-$ SuperKEKB collider~\cite{superkek}. It collects large data samples at the $\Upsilon(4S)$ resonance, where $B\bar{B}$ meson pairs are produced copiously. Belle operated from 1999 to 2010 and, together with the BaBar experiment, established the Kobayashi--Maskawa mechanism of $CP$ violation in the $B$-meson system, which led to the Nobel Prize in Physics in 2008~\cite{Bewan}. B factories produce $\tau^+\tau^-$ pairs, $c\bar{c}$ pairs, and $B\bar{B}$ meson pairs with comparable production cross sections and therefore also serve as charm and tau factories~\cite{Bewan}.  

Belle collected about $1~\mathrm{ab}^{-1}$ data  at the $\Upsilon(nS)$ ($n=1,2,3,4,5$) resonances~\cite{Belle}, while Belle II has accumulated over $900~\mathrm{fb}^{-1}$ of datasets~\cite{Belle2data}. These data-sets have been utilized to explore the possibility of  CLFV decays of $\tau$ leptons~\cite{taualp, Uno, tauto3l, tautolks, tauto3mu}, $B$ mesons~\cite{Btokstaul, Btoksttaul}, and bottomonium states~\cite{Y2stollp, Y1stollp, chibjtoll}. Some latest results of these CLFV decays are discussed below.

\section{\boldmath{$\tau$} CLFV decays at Belle and Belle II}
At the $\Upsilon(4S)$ resonance, the production cross section of $e^+e^- \to \tau^+\tau^-$ is comparable to that of the $e^+e^- \to B\bar{B}$ process, making Belle and Belle II excellent $\tau$ factories. In the center-of-mass (CM) frame, $e^+e^- \to \tau^+\tau^-$ events produce two $\tau$ leptons with back-to-back momenta, and each $\tau$ lepton carries an energy approximately equal to the beam energy,  $E_{\tau}^* = E_{\rm beam}^*$.  The $\tau$ lepton signal is identified using $M_{\rm BC}$ and $\Delta E$ distributions that peak at $\tau$ mass and zero, respectively, as seen in Fig.~\ref{mug} in Section~\ref{tautomug}. The corresponding variables are defined as, 

\begin{equation}
M_{\rm BC} = \sqrt{(E_{\rm beam}^*)^2 - (|\vec{p}_{\tau}^*)^2}
\end{equation}

\begin{equation}
\Delta E = E_{\tau}^* - E_{\rm beam}^*
\end{equation}

\noindent where $\vec{p}_{\tau}$ is the momentum vector of reconstructed $\tau$ lepton.  The decay products of the two $\tau$ leptons are well separated and can be assigned to opposite hemispheres using the thrust axis, allowing a clean separation of 1-prong and 3-prong decays. Consequently, $\tau$ decays provide an ideal environment for precision tests of the SM and searches for CLFV~\cite{Brignole, Calibbi, Ellis}.

\subsection{Invisible decays of axion-like particle $\alpha$ in $\tau^- \to \ell^- \alpha$ $(\ell=e, \mu$)}  
\begin{figure}[htb]
    \begin{center}
        {\includegraphics[scale=0.4]{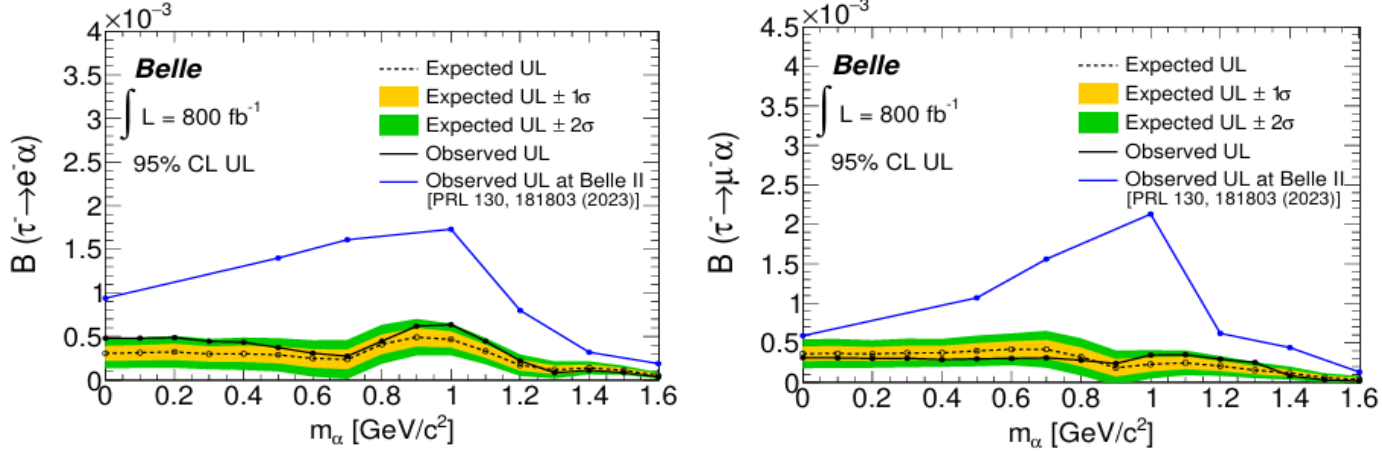}}\hspace{0.1cm}
             \caption{The $90\%$ CL upper limits on $\mathcal{B}(\tau^- \to e^- \alpha)$ (\it left) and $\mathcal{B}(\tau^- \to \mu \alpha)$  (\it right), including the expected 1 and 2 $\sigma$ yellow and green bands, respectively. The dashed (solid) curves are expected (observed) upper limits.
The blue curve shows the observed upper limits at Belle II~\cite{taualp}.}
\label{alpha}
    \end{center}
\end{figure}

An axion-like particle (ALP), denoted by $\alpha$, shares the quantum numbers of the QCD axion but is not constrained by its mass-coupling relation. ALPs arise naturally in a wide range of extensions of the SM and can mediate interactions between dark matter and SM particles. When ALPs couple to charged leptons, they can have CLFV couplings through $\tau^- \to \ell^- \alpha$. A previous search for this decay was performed by the Belle II experiment using a data sample of 62.8 fb$^{-1}$~\cite{taualp}. The Belle experiment has recently updated the results of this search using the 800 fb$^{-1}$ data sample by using both the $1\times1$ and $3\times1$ prong topologies~\cite{Uno}. In these topologies, one $\tau$ candidate is tagged using a 1-prong or 3-prong hadronic decay, while the signal-side $\tau$ is required to decay into a single charged lepton and an invisible ALP, resulting in missing momentum. The dominant background arises from the $\tau \to \ell \nu_{\ell} \bar{\nu}_{\tau}$ process. A search for a narrow ALP resonance is performed by fitting the lepton momentum distribution. No significant signal is observed, and improved $95\%$ confidence-level (CL) upper limits on the branching fractions of $\tau \to \ell \alpha$ are set for ALP masses in the range $0 < m_{\alpha} < 1.6$ GeV/$c^2$, as shown in Fig.~\ref{alpha}. Belle II is expected to update its results on this topic in the near future using a new method discussed in Ref.~\cite{new_method}.

\subsection{Search for CLFV decay $\tau \to \mu \gamma$}
\label{tautomug}
\begin{figure}[htb]
    \begin{center}
        {\includegraphics[scale=0.25]{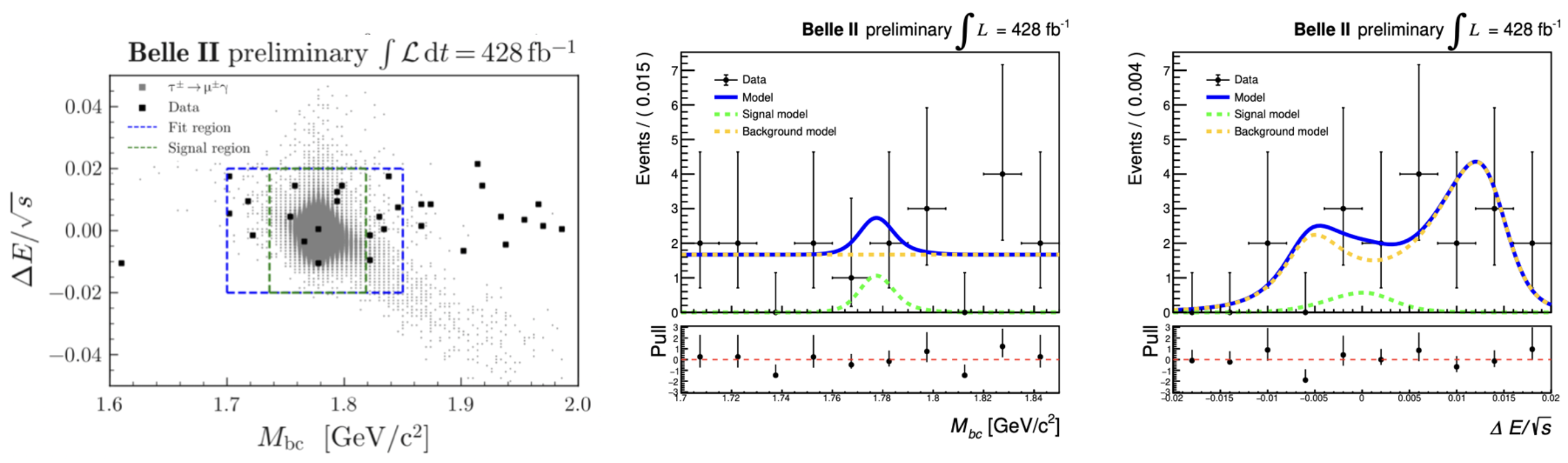}}\hspace{0.1cm}
 
 \caption{The $M_{\rm BC}$ versus $\Delta E/\sqrt{s}$ distribution for the $\tau^- \to \mu^-\gamma$ search (left), and its projections onto $M_{\rm BC}$ (middle) and $\Delta E/\sqrt{s}$ (right), together with their pull distributions. The black points represent data, while the blue solid curves denote the total fit. The green dotted and orange dashed curves represent the signal and background components, respectively. }
 
        \label{mug}
    \end{center}
\end{figure}

The new physics theories predict the branching fraction of $\tau^- \to \mu^- \gamma$ of the order of $10^{-10} - 10^{-8}$~\cite{Ilakovac}. Current best limit on $\mathcal{B} (\tau^- \to \mu \gamma)$ to be less than  $4.2 \times 10^{-8}$ comes from $988$ fb$^{-1}$ of the Belle data. Belle II has recently updated the results of this decay with 428 fb$^{-1}$ data using $1\times 1$ event topology~\cite{tautolg}. The signal is reconstructed with $\tau^- \to \mu^- \gamma$ and  tag side $\tau$ decays to 1-prong inclusive with muon veto. Belle II has used a Boosted Decision Tree (BDT) classifier for the first time to achieve $50\%$ increased signal efficiency with $80\%$ decrease in background rate. The search for the $\tau^- \to \mu^- \gamma$ signal is performed in two-dimensional (2D) plane of $M_{\rm BC}$ and $\Delta E/\sqrt{s}$ distributions, as shown in Fig.~\ref{mug} (left),  where $\sqrt{s}$ is the CM energy.  The sidebands data of these variable are used to derive the expected background in the signal region.  After unblinding the full data sample, 19 events are observed which are consistent with the expected background events of $15.7 \pm 3.4$. The obtained signal yield from 2D fit to the $M_{\rm BC}$ and $\Delta E/\sqrt{s}$ distributions is consistent with zero, as seen in Fig.~\ref{mug} (middle and right), and hence  the null results are reported. The $90\%$ CL upper limits on  are branching fractions are set to be $\mathcal{B}(\tau^- \to \mu^- \gamma)^{\rm obs} < 9.5 \times 10^{-8}$, which is consistent  with the expected limit   $\mathcal{B}(\tau^- \to \mu^- \gamma)^{\rm  exp} < 5.8 \times 10^{-8}$~\cite{tautolg}.

\subsection{Search for CLFV $\tau^- \to e^{\mp} \ell^{\pm}  \ell^-$}
The $\tau^- \to \ell^{\mp} \ell^{\pm} \ell^-$ decays provide a sensitive probe of CLFV, with the corresponding final states involving electrons and muons. The decay rates of these modes can be enhanced in models such as the type-II seesaw model and can also provide a probe of ALPs through $\tau^- \to X\ell^-$, followed by $X \to \ell^+\ell^-$~\cite{tauto3l}. The current existing limits on these decays, which range from $1.5$ to $2.7 \times 10^{-8}$, are based on $782~\mathrm{fb}^{-1}$ of Belle data. Belle II has recently performed this search using $428~\mathrm{fb}^{-1}$ of data with an inclusive-tagging method, similar to that used in the search for the $3\mu$ final state~\cite{tauto3l}.

The search for signal events is performed in the two-dimensional plane of the $\ell\ell\ell$ invariant mass ($M_{\ell\ell\ell}$) and $\Delta E_{\ell\ell\ell}$, which are expected to peak near the $\tau$ mass and zero, respectively. A BDT classifier is used to suppress the background. The signal yield is extracted by performing an unbinned one-dimensional maximum-likelihood (ML) fit to the $M_{\ell\ell\ell}$ distribution in the signal region defined by $-0.3 < \Delta E_{\ell\ell\ell} < 0.1$~GeV and $1.4 < M_{\ell\ell\ell} < 2.0$~GeV/$c^2$. The signal probability density function (PDF) is described by an asymmetric double-sided Crystal Ball (CB) function, while the background PDF is described by an exponential function. The expected number of background events, $N_{\rm exp}$, is obtained from a fit to the sideband regions. No significant signal is observed, and the $90\%$ CL upper limits on the branching fractions are determined to be:
\begin{itemize}
\item $\mathcal{B}(\tau^- \to e^- e^+ e^-) < 2.7 \times 10^{-8}$ (expected) and $< 2.5 \times 10^{-8}$ (observed);
\item $\mathcal{B}(\tau^- \to e^- e^+ \mu^-) < 2.1 \times 10^{-8}$ (expected) and $< 1.6 \times 10^{-8}$ (observed);
\item $\mathcal{B}(\tau^- \to e^- \mu^+ e^-) < 1.7 \times 10^{-8}$ (expected) and $< 1.6 \times 10^{-8}$ (observed);
\item $\mathcal{B}(\tau^- \to \mu^- \mu^+ e^-) < 1.6 \times 10^{-8}$ (expected) and $< 2.4 \times 10^{-8}$ (observed);
\item $\mathcal{B}(\tau^- \to \mu^- e^+ \mu^-) < 1.4 \times 10^{-8}$ (expected) and $< 1.3 \times 10^{-8}$ (observed).
\end{itemize}
\noindent Obtained limits are the most stringent to date for all modes except $\tau^- \to e^- \mu^+e^-$.

\subsection{Search for CLFV $\tau^- \to \ell^- \eta$}  The CLFV $\tau^- \to \ell^- \eta$ may include the new physics contributions from Leptoquark, Little Higgs with T-parity and type I-II Seesaw models~\cite{tautoleta}. Previous best limits on $\mathcal{B}(\tau^- \to e^- \eta) < 9.2 \times 10^{-8}$ and  $\mathcal{B}(\tau^- \to \mu^- \gamma) < 9.2 \times 10^{-8}$ were achieved using 427.9 fb$^{-1}$ data of Belle experiment~\cite{tautoleta}.  This measurement has recently updated with 427.9 fb$^{-1}$ data by performing 4 orthogonal searches in which tag side $\tau$ is based on 1-prong $\tau$ decay and signal side $\tau$ is reconstructed with either electron or muon and $\eta$ with its decay modes of $\eta \to \gamma \gamma$ and $\eta \to \pi^+\pi^-\pi^0$.  The dominant background from $\eta \to \gamma \gamma$ has been suppressed by a BDT classifier and after that it is estimated by a counting method. Whereas the  $\eta \to \pi^+\pi^-\pi^0$  is a clean decay mode and its background has been estimated by performing a 2D fit to the $M_{\tau}$ and $\Delta E$ distributions. No significant signal is observed in any of the $\eta$ decay mode. The $90\%$ CL upper limits are set to be $\mathcal{B}(\tau^- \to e^- \eta) < 9.21 \times 10^{-8}$ and $\mathcal{B}(\tau \to \mu^- \eta) < 4.23 \times 10^{-8}$.  The obtained limit on $\mu$ channel is most stringent to date and limit on $e$ channel is compatible with previous results~\cite{tautoleta}. 

\section{Search for CLFV  B decays}
Recent experimental anomalies in $b$-transitions, such as measurements of $R(D)$, $R(D^*)$ and $B^+ \to K^+\nu\bar{\nu}$, challenge the completeness of the SM~\cite{HFLAV, Allwicher}. The measured branching fraction of $B^+ \to K^+ \nu \bar{\nu}$ is about $2.7 \sigma$  from the SM prediction~\cite{Allwicher}. These anomalies may hint for new physics beyond the SM. Starting from the $B^+ \to K^+\nu\bar{\nu}$, theory predicts an enhancement of the branching fraction of $B \to K \tau^{\pm} \mu^{\mp}$ to be up to the level of $[2, 3] \times 10^{-6}$~\cite{Allwicher}, which is close to the experimental sensitivity.  
 Many CLFV decay modes have already been explored and reported null results~\cite{HFLAV}. Among them the measurements related to $b \to s \tau l$ and $b \to d \tau l$ are still less precise among the other measurements~\cite{HFLAV}.  
 
 At Belle and Belle II, a pair of $B$-mesons is produced at $\Upsilon(4S)$ resonance. One of the $B$ mesons is tagged with  either hadronic decay modes or semi-leptonic decays or inclusively decays and other $B$ is allowed to decay the signal mode of interest.  The purity hadronic tag is high but it has less efficiency, order of $0.5\%$. On the other hand, the efficiency of inclusive tag is almost $100\%$, but purity is less than $1\%$. A Full Event Interpretation (FEI) based on multivariate algorithm has been developed for the tag-side $B$ meson ($B_{\rm tag}$) with high efficiency~\cite{Btokstaul}. 
 
 \subsection{Search for the CLFV decay $B^0 \to K_S^0\tau^{\pm}\ell^{\mp}$}

This search uses the full Belle data sample of 711~fb$^{-1}$ together with 365~fb$^{-1}$ of Belle II data collected at the $\Upsilon(4S)$ resonance~\cite{Btokstaul}. One $B$ meson is reconstructed in hadronic decay modes using the FEI algorithm, while the remaining tracks are used to reconstruct the signal-side decay $B^0 \to K_S^0\tau^{\pm}\ell^{\mp}$ ($\ell=e,\mu$). The channels $B^0 \to K_S^0\tau^- \ell^+$ and $B^0 \to K_S^0\tau^+\ell^-$ are referred to as same-sign ($SS_\ell$) and opposite-sign ($OS_\ell$) modes, respectively. A BDT classifier is employed to optimize the event selection criteria. The $B_{\rm tag}$ reconstruction efficiency is calibrated using the control sample $B^0 \to D^-\pi^+$, while the signal PDF and BDT response are calibrated using $B^0 \to D_s^+D^-$ decays.

\begin{figure}[htb]
    \begin{center}
        {\includegraphics[scale=0.4]{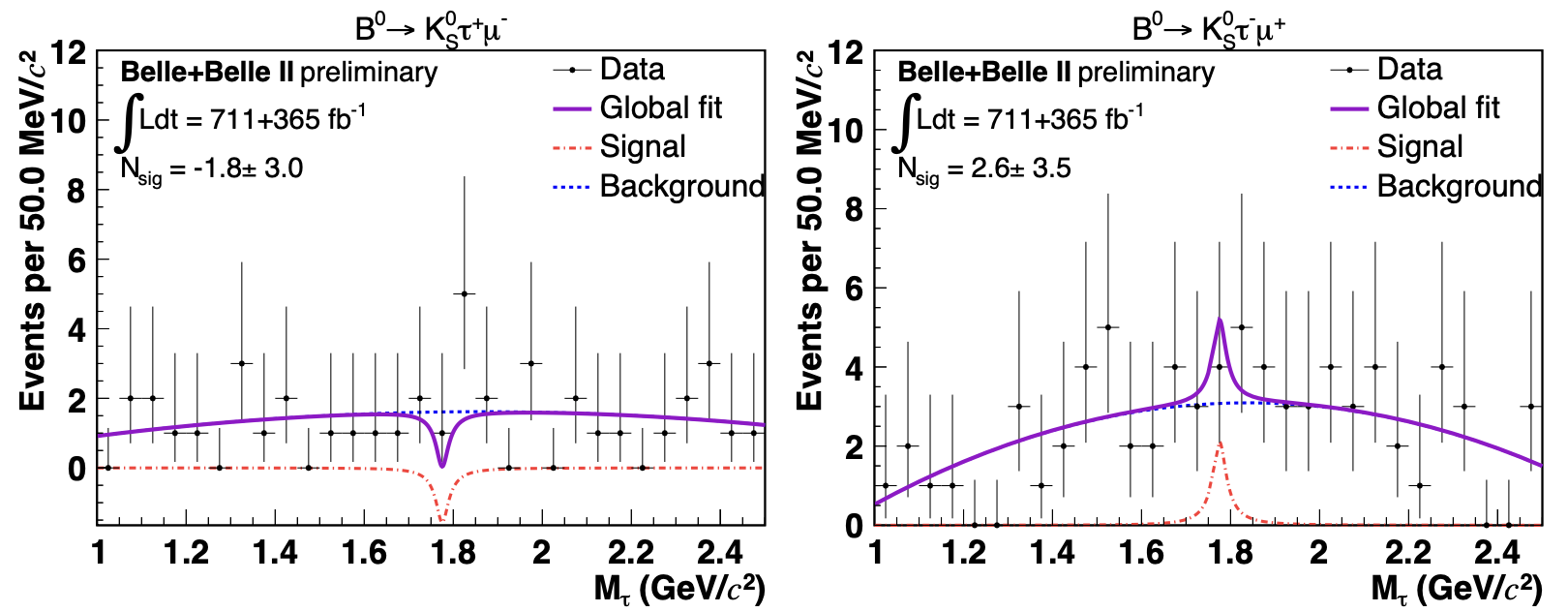}}\hspace{0.1cm}
 
 \caption{Fits to the $M_{\tau}$ distributions of $B^0 \to K_S^0 \tau^{\pm}\ell^{\mp}$ for the combined Belle and Belle II datasets. The black dots with error bars represent the data; the red dash-dotted, blue dashed, and purple solid curves represent the signal, background, and total fit, respectively.}
 
        \label{fig:LFVFit}
    \end{center}
\end{figure}

The signal yield is extracted from an unbinned ML fit to the reconstructed $\tau$-mass ($M_\tau$) distribution. Figure~\ref{fig:LFVFit} shows the fit results for the $B^0 \to K_S^0 \tau^{\pm} \mu^{\mp}$ as example. No statistically significant signal is observed in any channel. Consequently, the first 90\% CL upper limits on the branching fractions of $B^0 \to K_S^0\tau^{\pm}\ell^{\mp}$ are established:
\begin{align*}
\mathcal{B}(B^0\to K_S^0\tau^+\mu^-) &< 1.1\times10^{-5},  & 
\mathcal{B}(B^0\to K_S^0\tau^-\mu^+) < 3.6\times10^{-5}\\
\mathcal{B}(B^0\to K_S^0\tau^+e^-) &< 1.5\times10^{-5},  &
\mathcal{B}(B^0\to K_S^0\tau^-e^+)  < 0.8\times10^{-5}
\end{align*}
representing the first experimental limits on these lepton-flavor-violating decay modes.

\subsection{Search for the CLFV decay $B^0 \to K^{*0} \tau^{\pm}\ell^{\mp}$}
The combined data samples from Belle and Belle II, corresponding to integrated luminosities of 711 fb$^{-1}$ and 365 fb$^{-1}$, respectively, are also used to search for the decay $B^0 \to K^{*0} \tau^{\pm}\ell^{\mp}$  with $K^{*0} \to K\pi$~\cite{Btoksttaul}. Previously, LHCb searched for this decay mode using a 9 fb$^{-1}$ data set and obtained upper limits of $0.82~(1.0)\times 10^{-5}$ at $90\%$ CL.
Four different final states are distinguished according to the flavor of the final-state lepton $\ell$ and the sign of its charge relative to that of the kaon from the $K^{*0}$: $SS_{\ell}$ for $B^0 \to K^{*0}(\to K^+\pi^-) \tau^- \ell^+$ and  $OS_{\ell}$ for $B^0 \to K^{*0}(\to K^+\pi^-) \tau^+\ell^-$. The same method for optimizing the event selection criteria and extracting the signal yield as that used for $B^0 \to K_S^0 \tau^{\pm}\ell^{\mp}$~\cite{Btokstaul} is applied.  No significant signal is observed. The $90\%$ CL upper limits on the branching fractions are calculated as follows:
\begin{align*}
\mathcal{B}(B^0\to K^{*0} \tau^+\mu^-) &< 4.2 \times10^{-5}, &
\mathcal{B}(B^0\to K^{*0} \tau^-\mu^+) &< 5.6\times10^{-5}, \\
\mathcal{B}(B^0\to K^{*0} \tau^+e^-) &< 2.9 \times10^{-5}, &
\mathcal{B}(B^0\to K^{0} \tau^-e^+) &< 6.4\times10^{-5}.
\end{align*}
Among these, the upper limits for the electron modes are the most stringent to date.

\section{Search for CLFV in Bottomonium Decays}

The CLFV decays of bottomonium can provide useful information about several classes of operators, such as vector, axial-vector, and tensor operators, involved in four-fermion interactions~\cite{Hazard}. Precise measurements of two-body CLFV decays of vector mesons allow one to effectively probe vector and tensor operators~\cite{Hazard}. Belle has recently set one of the most stringent limits on CLFV decays $\Upsilon(2S) \to \ell\tau$ ($\ell = e, \mu$) using 158 million $\Upsilon(2S)$ events~\cite{Y2stollp}. The same $\Upsilon(2S)$ dataset has also been utilized to report first experimental limits on  two-body CLFV decays $\Upsilon(1S) \to \ell\ell'$ ($\ell \ne \ell'$, $\ell\ell' = e,\mu,\tau$) and three-body CLFV decays $\Upsilon(1S) \to \gamma\ell\ell'$~\cite{Y1stollp}. Here, the $\Upsilon(1S)$ sample is selected by tagging the pion pair from the $\Upsilon(2S) \to \pi^+\pi^- \Upsilon(1S)$ transition. The reported limits on the three-body CLFV decays $\Upsilon(1S) \to \gamma\ell\ell'$ can be used to constrain the corresponding Wilson coefficients of axial-vector, scalar, and pseudoscalar operators~\cite{Hazard}.

\subsection{CLFV decays of $\chi_{bJ}(1P)$}
CLFV decays of $\chi_{b_0}(1P)$ provide a low-energy probe of CLFV in the scalar sector~\cite{Hazard}, complementary to direct searches for CLFV in Higgs boson decays. CLFV decays of $\chi_{b_1}(1P)$ and $\chi_{b_2}(1P)$ probe axial-vector and tensor operators, respectively. Searches for CLFV decays of $\chi_{b_J}(1P)$ are performed for the first time using a sample of $(158 \pm 4) \times 10^6$ $\Upsilon(2S)$ mesons, from which $\chi_{b_J}(1P)$ candidates are selected through the transition $\Upsilon(2S) \to \gamma_1\chi_{b_J}(1P)$~\cite{chibjtoll}. The cascade decays $\chi_{b_J}(1P) \to \gamma_2\Upsilon(1S)$, followed by $\Upsilon(1S) \to e^+e^-$ or $\Upsilon(1S) \to \mu^+\mu^-$, are also studied as control modes to characterize background rates, estimate systematic uncertainties, and validate the analysis procedure. The measured branching fractions of the control modes are consistent with the corresponding Particle Data Group values within $(10\text{--}20)\%$~\cite{chibjtoll}.

\begin{figure}[htb]
    \begin{center}
        {\includegraphics[scale=0.28]{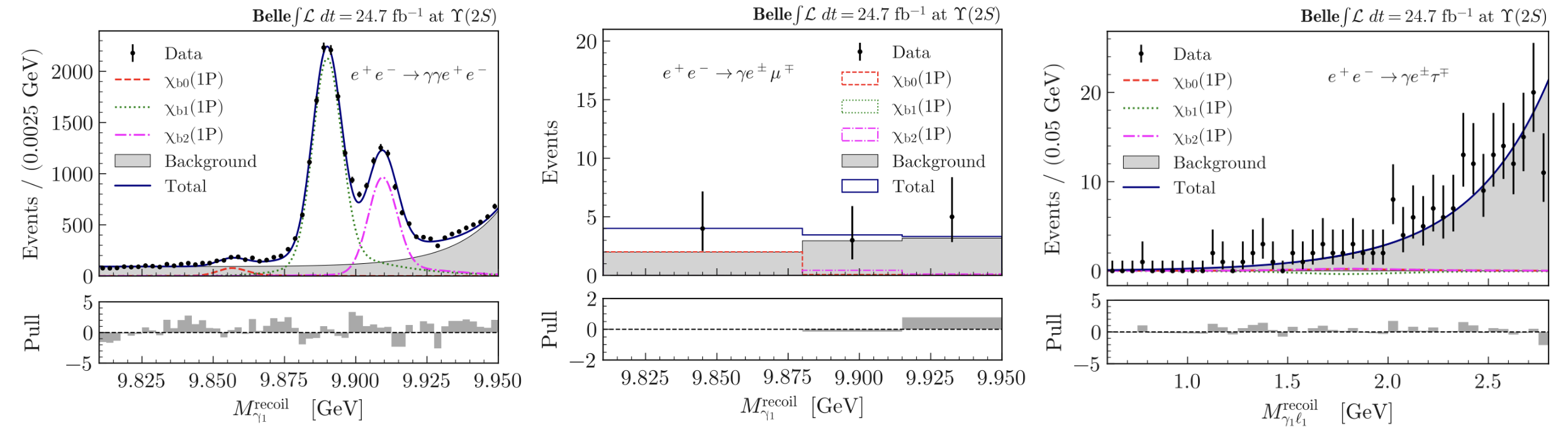}}\hspace{0.1cm}
 
 \caption{Examples of fits to the recoil mass of the $\gamma_1$ system for the control mode $\chi_{b_J} \to \gamma \Upsilon(1S) (\to e^+e^-)$ (left) and the CLFV $\chi_{b_J} \to e\mu$ decay (middle), as well as to the recoil mass of the  $\gamma_1\ell_1$ system for $\chi_{b_J} \to e \tau$ (right). Black points with error bars represent the data; the grey filled histogram shows the background contribution. The dashed red, green, and cyan curves correspond to the  $\chi_{b_0}$, $\chi_{b_1}$ and $\chi_{b_2}$ signal components, respectively,  while the solid black curve shows the total fit.}
 
        \label{chibjdecays}
    \end{center}
\end{figure}

The signal yields are extracted from fits to the recoil mass of the $\gamma_1$ system, $m_{\gamma_1}^{\rm recoil}$, for both  control modes and the CLFV decays $\chi_{b_J} \to e^{\pm}\mu^{\mp}$, and to the recoil mass of the $\gamma_1\ell_1$ system, $m_{\gamma_1\ell_1}^{\rm recoil}$, for $\chi_{b_J} \to \ell_1^{\pm}\tau^{\mp}$. Examples of fits to the recoil mass of the $\gamma_1$ system for the control mode and the $\chi_{b_J} \to e\mu$ decay, as well as to the recoil mass of the  $\gamma_1\ell_1$ system, are shown in Fig.~\ref{chibjdecays}. The obtained signal yields for all the analyzed CLFV decays $\chi_{b_J} \to \ell\ell'$ are consistent with zero. The $90\%$ CL upper limits on the branching fractions are determined using \texttt{pyhf} for $\chi_{b_J}(1P) \to e^{\pm}\mu^{\mp}$ and a frequentist calculator from  \texttt{RooStats} for $\chi_{b_J} \to \ell^{\pm}\tau^{\mp}$ decays, including systematic uncertainties. The measured $90\%$ CL upper limits on the branching fractions of the CLFV decays $\chi_{b_0}(1P) \to \ell\ell'$ are translated into constraints on the previously unexplored Wilson coefficients of the left- and right-handed scalar operators, $C_{\rm SL}^{q\ell_1\ell_2}$ and $C_{\rm SR}^{q\ell_1\ell_2}$, which can constrain the parameter space of the NP models.

\section{Summary}  
CLFV decays are sensitive probes of new physics beyond the SM. The large data samples collected by Belle and Belle II have been utilized to search for CLFV decays in $\tau$, $B$, and bottomonium decays. No significant signal is observed in the searches for any CLFV decays, and some of the most stringent upper limits on these decay modes have been set. More results are expected in the near future, particularly with the future Belle II data sample.

\section{Acknowledgements}
This work is supported  by the Seed Funding of Jilin University.

\end{document}